\documentclass{iopart}
\usepackage{graphicx}
\usepackage{bm}

\begin{document}
\title{Linear Algebra and Charge Self-consistent Tight-binding Method 
for Large-scale Electronic Structure Calculations}
\author{T. Fujiwara$^{1,2}$, S. Yamamoto$^{3,2}$, T. Hoshi$^{4,2}$, S. Nishino$^{1,2}$, H.Teng$^{1,2}$, 
T. Sogabe$^{5,2}$, S.-L. Zhang$^{6,2}$, M. Ikeda$^7$, M. Nakashima$^8$, and N. Watanabe$^9$} 
\address{(1) Center for Research and Development of Higher Education, 
The University of Tokyo, 
Bunkyo-ku, Tokyo, 113-8656, Japan}
\address{(2) Core Research for Evolutional Science and Technology, 
Japan Science and Technology Agency (CREST-JST), Japan}
\address{(3) School of Computer Science, Tokyo University of Technology, Katakura-machi, Hachioji, Tokyo 192-0982, Japan} 
\address{(4) Department of Applied Mathematics and Physics, Tottori University, Tottori 680-8550, Japan}
\address{(5) School of Information Science and Technology, Aichi Prefecture University, Nagakute-cho, Aichi 480-1198, Japan}
\address{(6) Department of Computational Science and Engineering, Nagoya University, Chikusa-ku, Nagoya 464-8603, Japan}
\address{(7) Nanoelectronics Research Center, Fujitsu Laboratories LTD., Atsugi 243-0197, Japan}
\address{(8) Semiconductor Energy Laboratory Co. LTD., Atsugi, Kanagawa 243-0036, Japan}
\address{(9) Mizuho Information \& Research Institute Inc., Chiyoda-ku, Tokyo 101-8443, Japan}

\begin{abstract}
We review our recently developed electronic structure calculation methods 
for the dynamics of large-scale solids, liquids or soft materials   
with an efficient algorithm  of linear algebra. 
The Hamiltonian of electronic structures is based on 
the `atomic superposition and electron delocalization molecular orbitals theory' (ASED), 
using the Mulliken charge density. 
Very crucial algorithms of the linear algebra are 
the generalized Lanczos method and 
the generalized shifted COCG (conjugate orthogonal conjugate gradient) method 
for general eigen-value problems.   
The shift equation and the seed switching method are essential for the shifted COCG method 
and reduce the computational cost much. 
We, then, present some applications to electronic structure calculations with MD simulation, 
such as the fracture propagation in nano-scale Si crystals and the proton transfer in water. 
\end{abstract}

\date{\today}
\maketitle

\section{Introduction}

The electronic structure calculations in nano-scale structures have attracted 
considerable attention. 
It is much desirable to achieve molecular dynamics (MD) simulations 
with quantum mechanical electronic structure calculation for systems of several hundred thousands 
atoms with a few hundred pico-seconds (or longer time) process, 
since the atomic structure is very essential to the electronic structure and {\it vice versa}.  
Requirement for higher accuracy of the total electronic structure energy in nano-scale systems 
and that for larger system size contradict each other. 
One of possible choices for satisfying these requirements could be 
the LDA (local density approximation) calculation with massive-parallel computation. 
Even in that case, a long-time process for MD simulations is another difficulty.

In the present paper, we propose one possible way to pursue a quantum mechanical MD simulation 
of large system size and long time process 
at the expense of lower accuracy of the electronic structure energy. 
The first-principles tight-binding method is one of promising techniques and 
we review the ``{\it atom-superposition and electron-delocalization molecular-orbitals theory}'' with 
self-consistent charge density in Section \ref{ASED}.
Then, based on the tight-binding formulation, 
we develop novel algorithm of linear algebra for large scale systems 
with the non-orthogonal local orbitals in Section \ref{Krylov}. 
Section \ref{Application} is devoted to present applications to nano-scale systems, 
the fracture propagation and surface reconstruction in Si single crystals and 
the proton transfer in water. 
Section \ref{Conclusion} is conclusions.

\section{First-principles tight-binding methods and charge self-consistency}

\subsection{Several first-principles tight-binding methods}

We have several semi-empirical tight-binding formalisms~\cite{Kwan1994,Vogl-1983}
whose results are comparable to those of the first-principles LDA calculation. 
We have also several theories of first principles tight-binding Hamiltonian, e.g. 
the tight-binding LMTO method,~\cite{tb-LMTO} 
the LDA tight-binding method,~\cite{LDA-tb} 
the first principles tight-binding method of the Naval Research Laboratory (NRL).~\cite{tb-NRL} 
We explain in the next subsection the tight-binding Hamiltonian of 
the `atom-superposition and electron-delocalization molecular-orbital' theory.~\cite{ASED}  

\subsection{Atom-superposition and electron-delocalization molecular-orbital theory}\label{ASED}

The {\it atom-superposition and electron-delocalization molecular-orbital} (ASED) theory 
is based on a charge-density partitioning method, 
where the charge density is determined by the Mulliken method of partitioning the overlap charge density 
(the Mulliken charge) of assumed atomic orbitals.

Diagonal elements of the Hamiltonian matrix in the extended H\"uckel theory  
would be given as the valence orbital ionization potential with a quadratic formula of the Mulliken charge.  
In the present form of the ASED theory, a diagonal element of the Hamiltonian $H_{i\alpha,i\alpha}$, where $i$ and $\alpha$ 
denote an atom and its orbital,  
is set to equal to 
an atomic energy level (ionization energy) $\epsilon_{i\alpha}$.
An off-diagonal element of the Hamiltonian is determined by the bond formation and 
an extended H\"uckel-like delocalization energy may be generally a good approximation;~\cite{general-Hueckel} 
\begin{eqnarray}
H_{i\alpha,j\beta}=K(\alpha, \beta; R_{ij})\frac{\{H_{i\alpha,i\alpha}+H_{j\beta,j\beta}\}}{2}S_{i\alpha j\beta} ,
\label{Hap-1}
\end{eqnarray}
where  a coefficient $K(\alpha, \beta; R_{ij})$ stands for distortion of wavefunctions 
depending upon orbitals and a distance $R_{ij}$,~\cite{Calzaferri-1996}
and $S_{i\alpha,j\beta}$ is the overlap integral.

Another contribution to the total energy is the ion-core repulsive energy,  
which is defined, in the tight-binding density functional (DFT) formalism~\cite{tb-DFT}, 
to be the difference between LDA energy and the tight-binding band structure energy.

The charge transfer effects is crucial within the charge self-consistent (CSC) theory 
and formulated by the second order perturbation theory,~\cite{Elstner} 
which gives a form of the extended Hubbard-type Coulomb interaction as
\begin{eqnarray}
 H_{i\alpha,j\beta}^{\rm CSC} = \frac{1}{2}S_{i\alpha,j\beta}\sum_{k} (\gamma_{ik}+\gamma_{jk}) \Delta q_k
\label{Hap-CSC}
\end{eqnarray}
where $\Delta q_k$ is the deviation of the (Mulliken-type) charge of an atom $k$ 
and $\gamma_{ik}$ is the (diagonal or off-diagonal) Coulomb interaction between atoms 
$i$ and $k$ or the chemical hardness.

We are applying the CSC tight-binding method to water, molecular liquids and 
soft materials of nano-scale size ($\simeq 3000$~atoms).  
A MD simulation  of a proton transfer in water will be presented preliminarily in this paper.   

\section{Linear algebra and generalized Krylov subspace method for generalized eigen-value problem}\label{Krylov}
\subsection{Linear equation and Green's function~{\rm \cite{fujiwara-2008-2009}}}

In order to study electronic structures in nano-scale materials, it is much desirable 
to construct the order-$N$ algorithm whose computational load, e.g. cpu time and memory size, 
is proportional to the system size or the number of atoms.  

Problems on electronic properties in materials starts usually from obtaining  
eigen-energies and eigen-functions in materials 
or solving linear equations to find $| x^{(j)} \rangle$ 
for a given state $|j\rangle $ and a given energy $E$;
\begin{eqnarray}
     && (E-{\hat{\cal H}}) | x^{(j)} \rangle = |j\rangle ,
\label{base-2} 
\end{eqnarray}
where ${\hat{\cal H}}$ is a one-electron (Hermitian) Hamiltonian operator. 
Once we solve Eq.(\ref{base-2}) and find a vector $| x^{(j)} \rangle$, 
the matrix elements of the Green's function
$ \hat{G}(z)= (z-{\hat{\cal H}})^{-1} $ 
can be evaluated very easily as 
\begin{eqnarray} 
     && G_{ij}(E)= \langle i| (E+i\delta-{\hat{\cal H}})^{-1}|j\rangle =\langle i|x^{(j)} \rangle  
\label{GreensFunc} ,
\end{eqnarray}
where $\delta$ is an infinitesimally small (positive) number.  

\subsection{Generalized Lanczos method~~{\rm \cite{Teng-2010}}}
In case of the orthogonal basis set, 
the Lanczos process and 
the shifted conjugate orthogonal conjugate gradient (COCG) method~\cite{Vost-Melissen1990,FROMMER2003,TAKAYAMA2006} 
can be constructed in the framework of the Krylov subspace 
generated by ${\cal H}$ and ${\bm x}^{(0)}$.

When the basis set $\{\phi_i\}$ is non-orthogonal, the overlap of wavefunctions $|\psi_\alpha\rangle$ 
($|\psi_\alpha\rangle =\sum_i |\phi_i\rangle u_i^{(\alpha)}$) 
is define by the ``$S$-product'' as
\begin{eqnarray}
 \langle \psi_\alpha | \psi_\beta \rangle  =      ({\bm u}^{(\alpha)}, {\bm u}^{(\beta)})  
                                           \equiv \sum_{ij} {u_i^{(\alpha)}}^*u_j^{(\beta)} S_{ij}  \label{eq(0.12)}
\end{eqnarray}
and $S_{ij}= \langle \phi_i | \phi_j \rangle$ is an element of the overlap matrix.

With a non-symmetric (non-Hermitian) matrix $H=S^{-1}{\cal H}$, 
we can get the three-term recursive equation 
or the generalized Lanczos process as
\begin{eqnarray}
H{\bm u}^n&=& a_n{\bm u}^n +b_{n+1}{\bm u}^{n+1} +b_n{\bm u}^{n-1} \label{eq(1.13)} 
\end{eqnarray}
for $n=0,1,\cdots$ and $b_0=0$ 
with the ``S-orthogonality" as 
\begin{eqnarray}
({\bm u}^{n+1}, {\bm u}^{m})=0 \ \ \ {\rm for}\ \ \ \ m=n, n-1,n-2,\cdots\cdots , 0 .
\label{eq(1.18)}
\end{eqnarray}
We can set $b_n$'s ($n=1,2,\cdots$) to be positive numbers. 
The generalized Krylov subspace is spanned by the vectors ${\cal H}^{m}{\bm x}^{(0)}=(S^{-1}H)^m{\bm x}^{(0)}$ and 
defined as 
\begin{eqnarray}
{\cal K}_{n}({H}, {\bm x}^{(0)})={\rm span} \{{\bm x}^{(0)},{H}{\bm x}^{(0)},{H}^2{\bm x}^{(0)},\dots, {H}^{n-1}{\bm x}^{(0)}\} ,
\label{eq:Krylov}
\end{eqnarray}

Within the set of basis $\{{\bm u}^{0},{\bm u}^{1},{\bm u}^{2},\cdots\}$ or 
the resultant `generalized' Krylov subspace, 
the `Hamiltonian matrix $H$' is tri-diagonalized on the basis of 
$|\psi_n\rangle = \sum_i | \phi_i \rangle u_i^{(n)}$.
Therefore, one can calculate the Green's function in this `generalized' Krylov subspace 
and also achieve the spectral decomposition (the subspace diagonalization) 
in the `generalized' Krylov subspace, which we call the generalized Lanczos method.

\subsection{Generalized shifted COCG method~{\rm \cite{GsCOCG,Teng-2010}}}

Even for non-orthogonal basis set, we can generalize the COCG procedure, 
similar to the original one,~\cite{TAKAYAMA2006} 
for the Hamiltonian ${\cal H}$ and the overlap matrix $S$. 
The linear equation of the `seed' energy (an arbitrary chosen energy) $\sigma_s$ can be written as 
\begin{eqnarray}
(S^{-1}{\cal H}+\sigma_s 1){\bm x}^{(i)} = S^{-1}{\bm b} \label{shift-E} .
\end{eqnarray}   
An approximate solution at the $n$-th iteration, 
a searching direction for the solution at the next iteration step, 
and a residual vector are written as ${\bm x}_n$, ${\bm p}_n$ and ${\bm r}_n$, respectively. 
We can get a recursive equations under an appropriate  initial conditions;
\begin{eqnarray}
 {\bm p}_n   &=& {\bm r}_{n}^\prime   + \beta_{n-1} {\bm p}_{n-1}            \label{Eq:CGp:p} , \\
{\bm x}_{n+1}&=& {\bm x}_{n} + \alpha_{n} {\bm p}_{n}                        \label{Eq:CGp:x} , \\
{\bm r}_{n+1}&=& {\bm r}_{n} - \alpha_{n} ( {\cal H}+\sigma_sS) {\bm p}_{n}   \label{Eq:CGp:r} ,
\end{eqnarray}
with
$
{\bm r}_{n+1}^\prime = S^{-1}{\bm r}_{n+1}  
$, 
$ \alpha_{n} = \frac{({\bm r}_{n}^\prime, {\bm r}_{n}^\prime)}{({\bm p}_{n}, ({H}+\sigma_s) {\bm p}_{n})} $ and
$ \beta_{n}  = \frac{({\bm r}_{n+1}^\prime,{\bm r}_{n+1}^\prime)}{({\bm r}_{n}^\prime,{\bm r}_{n}^\prime)}$.


For non-orthogonal basis set, we can generalize the shift equation.  
The linear equation of a `shift' energy $\sigma$ is as follows;
\begin{eqnarray}
(S^{-1}{\cal H}+\sigma 1){\bm x}^{(i)} = S^{-1}{\bm b} \label{shift-E} .
\end{eqnarray}

Then the entire information is delivered to the coefficients $\alpha_n^\sigma$ 
and $\beta_n^\sigma$ at every shift energy $\sigma$ 
by a scalar constant $\pi^{\sigma}_{n+1}$, which is generated by the recursive equation 
\begin{eqnarray}
	\pi^{\sigma}_{n+1}  &=& \{1+ \alpha_n (\sigma-\sigma_s) \} \pi^\sigma_n
                            - \frac{\beta_{n-1}}{\alpha_{n-1}}\alpha_{n}( \pi^\sigma_{n}- \pi^\sigma_{n-1})       . \label{Eq:shift:pi} 
\end{eqnarray}
The most important issue of the shift equation 
is the theorem of {\it collinear residual}:~\cite{FROMMER2003}
\begin{equation}
	{\bm r}^\sigma_{n}=\frac{1}{\pi^{\sigma}_{n}}{\bm r}_{n} .
\label{Eq:collinear}
\end{equation}
Then, once the solution is obtained at the seed energy by Eqs.~(\ref{Eq:CGp:p})$\sim$(\ref{Eq:CGp:r}),
solutions at arbitrary shift energies 
can be evaluated  by using only scalar-scaler and scalar-vector multiplications.

The only remaining problem is the calculation of $S^{-1}{\bm r}_n$. 
Fortunately, the overlap matrix $S$ is positive-definite and 
nearly equal to the unit matrix. 
Therefore, we can solve this equation efficiently by the CG method, e.g. 
solve inversely ${\bm r}_n =S{\bm r}_n^\prime$ for a given ${\bm r}_n$.

The choice of the seed energy $\sigma_s$ is not unique. 
If one would choose a seed energy in an energy range of faster convergence, 
spectra at majority energy points have not be converged yet after achieving 
the convergence at the seed energy.  
In that case, one might restart the calculation with new seed energy from the beginning.

The {\it seed-switching} is very efficient technique to avoid waste of the already finished 
part of constructing the Krylov subspace.~\cite{SOGABE2007,yamamoto2007b} 
One can choose a new seed energy $\sigma_s^{\rm new}$ 
and can continue the calculation without discarding the information of the previous calculation 
with the old seed energy $\sigma_s$. 

\begin{figure}[htbp] 
\begin{center}
\resizebox{0.7\textwidth}{!}{
  \includegraphics{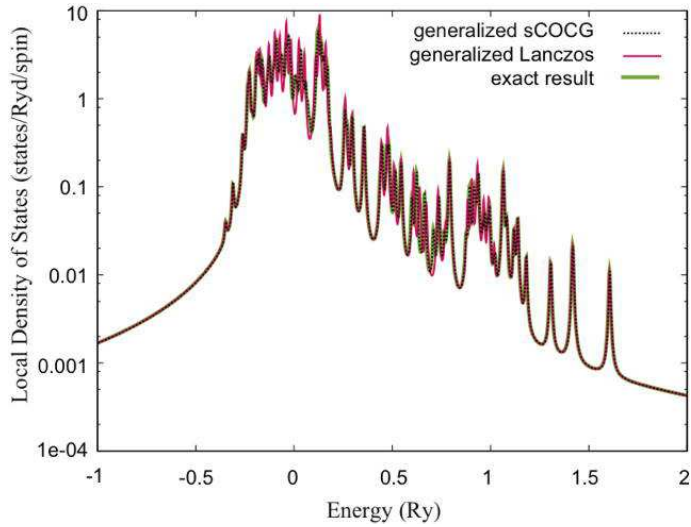}
}
\end{center}
\caption{\label{FIG-lDOS-test} 
The local density of states of fcc gold of Au~256 atoms by the tb-LDA with non-orthogonal basis set. 
There are three lines in the figure and they are almost overlapped. 
The chemical potential locates at $+0.2947$~Ry.
}
\end{figure}

\subsection{Example}

We have explained the generalized Lanczos method and the generalized shifted COCG method. 
Now we  present the calculated results of the local density of states (LDOS) 
by using the tight-binding Hamiltonian with  non-orthogonalized basis set 
given by the NRL Hamiltonian.~\cite{tb-NRL}
The system is of gold 256 atoms in a supercell of an fcc structure 
with s-, p-, and d-orbitals under a periodic boundary condition. 
Figure~\ref{FIG-lDOS-test} shows the LDOS of d-orbitals, where 
the vertical axis is in logarithmic scale to see the behavior of tails. 
The three calculated results, the generalized Lanczos method, the generalized shifted COCG method
and the exact calculation, are almost identical with each other.

The cpu times by a standard work-station of a single processor are 
3.92~s, 21.20~s and 57.6~s for the generalized Lanczos method, the generalized shifted COCG 
and the exact calculation, respectively.  
The algorithm of the present developed algebraic methods is suitable for parallel computation. 
on the other hands, the cpu time grows as a cubic power of the matrix size in the exact calculation. 
Therefore, the present methods are promising in large scale electronic structure calculations.


\section{Application to nano-scale systems}\label{Application}
\subsection{Scale-size dependence of recently developed linear algebraic methods}

\begin{figure}[thbp] 
\begin{center}
\resizebox{0.48\textwidth}{!}{
  \includegraphics{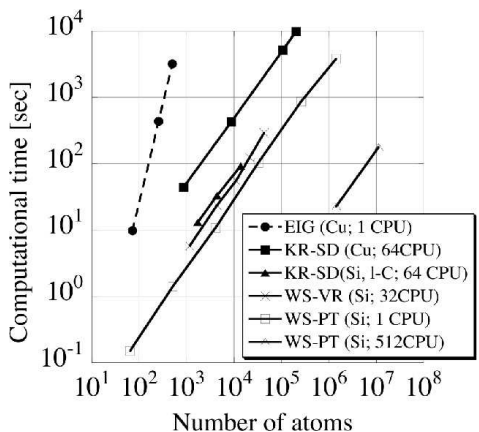}
}
\end{center}
\caption{\label{FIG-BENCH} 
The computational time as a function of the number of atoms ($N$). 
\cite{HOSHI2003, HOSHI2005, HOSHI2006}
The time was measured in a older standard workstation 
for metallic (fcc Cu and liquid C) and insulating (bulk Si) 
systems with up to 11,315,021 atoms, 
by the conventional eigenstate calculation (EIG) 
and by our methods for large systems; 
KR-SD (Krylov subspace-diagonalization), WS-VR (Wannier-state variational) and 
WS-PT (Wannier-state perturbation) methods.  
The CPU time increases linearly with increasing the number of atoms. 
See the original papers~\cite{HOSHI2003, HOSHI2005, HOSHI2006}
for the details of parallel computation.
}
\end{figure}

Much attention has been paid to nano-scale systems and 
the first principles molecular dynamics simulation has become more important 
in material sciences and engineering.
We have developed a set of computational methods of electronic structure calculations, 
{\it i.e.} the generalized Wannier state methods,~\cite{HOSHI2000,HOSHI2003,HOSHI2005,HOSHI2006} 
the Krylov subspace method,~\cite{TAKAYAMA2004} and the shifted COCG method for nano-scale systems.~\cite{TAKAYAMA2006} 
In the preceding sections, we have explained recent development on the ASED and generalized shifted COCG method. 
The tight-binding MD simulation method with being-developed linear algebraic algorithm 
is useful to achieve calculations in extra-large systems 
whose computational cost increases linearly proportional to the system size  
as shown in Fig.\ref{FIG-BENCH}, where tested examples are metallic, insulating and semiconducting 
dense materials.

\subsection{Fracture propagation and surface reconstruction in silicon crystal}\label{Si} 

Here, we will show a calculation results of fracture propagation 
in a bulk Si single crystal with the semi-empirical tight-binding Hamiltonian 
by Kwon {\it et al}~\cite{Kwan1994}, based on the orthogonal basis set. 
The cleavage propagation speed is estimated, in the present model, as 
$v_{\rm prop} \simeq 2~{\rm nm/ps} = 2~{\rm km/s}$, 
which is comparable to the velocity of 
the Rayleigh wave $v_{\rm R}=4.5~{\rm km/s}\simeq 4.5~{\rm nm/ps}$.~\cite{HOSHI2003,HOSHI2005} 

\begin{figure}[t] 
\begin{center}
\includegraphics[width=0.9\linewidth]{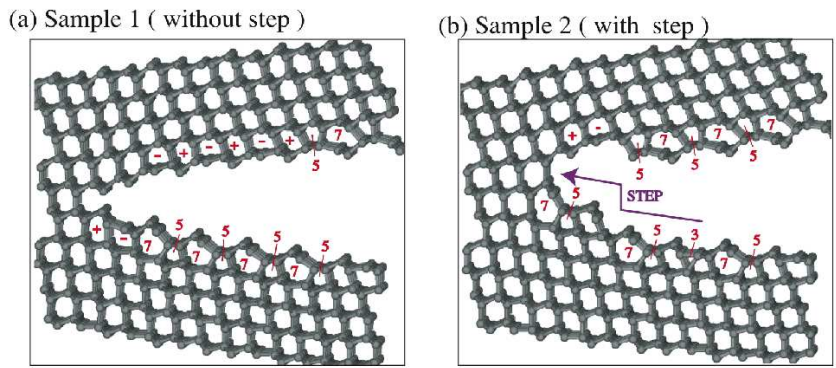}
\end{center}
\caption{ 
Fracture propagation on a (111) plane of a Si single crystal. 
(a) Appearance of Pandy structure without steps.
(b) Pandy structure with a step.
The numbers show the number of atoms in a ring. 
The symbols ($+$) or ($-$) 
refer to the electron charge, excess or deficiency.}
\label{Frac-Prop-111}
\end{figure}

The easy-propagating plane of fracture in Si is known widely to be that of (110) or (111) and 
we studied the phenomena of fracture propagation in 14~nm scale Si crystals.~\cite{HOSHI2005} 
In case of fracture on the (111) plane, the the Pandy structure~\cite{PANDEY} of 
(111)-$(2\times 1)$ surface reconstruction appears with several steps. 
The direction or the structure of fracture propagation planes is not explained 
by the stabilization energy of the final stable surfaces structure 
because a transient surface structure is first formed immediately after the bond-breaking 
and, after appearance of the ideal surface, 
the surface reconstruction happens to form the final reconstructed surface.  
In the process of fracture propagation and surface reconstruction, 
the two different kinds of energy loss and gain compete with each other, 
e.g. the energy competition between an electronic energy gain of breaking bonds 
and an elastic energy loss of distorting the lattice. 
This kind competition may be possible only in nano-scale systems 
and, in a MD simulation, we should prepare larger systems, e.g. a few tens nm length. 
Then, even a fracture propagation starts on a (001) plane, 
the plane of the fracture propagation changes to (111) and (110) planes. 
Figure \ref{Frac-Prop-111} shows examples of the surface reconstruction after 
formation of surfaces. 

\subsection{Proton transfer in water}\label{H3O-nH2O} 

The correlation energy of water dimer,~\cite{MCC-theory} hydrogen-bonded network~\cite{hydrogen-bond}
and protonated water~\cite{protonated-water} are all long-investigated problems 
of quantum chemistry. 
We have been applying the ASED method to water and proton transfer in water to investigate the applicability 
of the method.

\begin{figure}[t] 
\begin{center}
\includegraphics[width=0.9\linewidth]{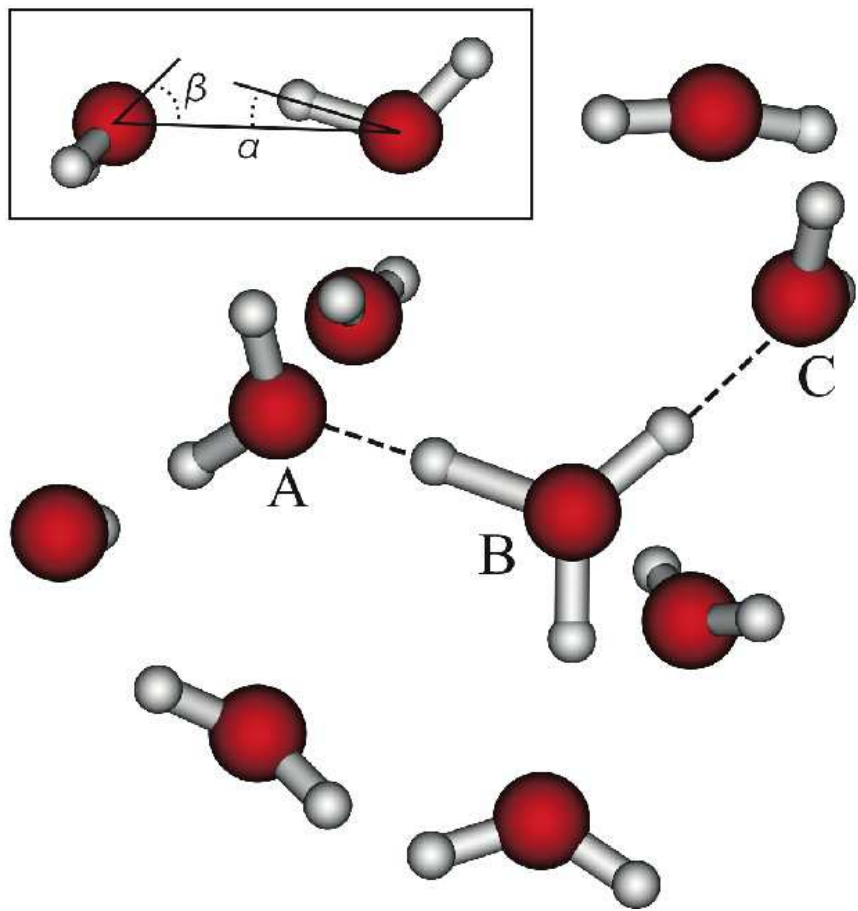}
\end{center}
\caption{ 
Snap shot of the proton transfer process. 
A proton transfers from a H$_2$O molecule A to that of B and   
another proton will transfer from the oxonium ion H$_3$O$^+$ to that of C. 
The inset shows the definition of angles $\alpha$ and $\beta$ for a stable configuration of 
a water dimer.}
\label{Proton}
\end{figure}

The bond angle H-O-H (an experimental value 104.5$^\circ$) is determined sensitively by 
the energy levels (binding energy and bond angle dependence) 1B$_2$ and 3A$_1$ 
of H$_2$O, which could be done by adjusting the ionic energy levels $\epsilon_{\rm 2s}$ and 
$\epsilon_{\rm 2p}$ of oxygen 2s and 2p orbitals and their STO parameters. 
After that,  we could obtain the correct order of molecular energy levels 
of (2A$_1$), (1B$_2$), (3A$_1$), (1B$_1$), (4A$_1$), (2B$_2$) and an approximate 
cohesive energy of a molecule (10.97~eV, Cf. the experimental value of 10.08~eV). 
The bond length can be adjusted by slight rescaling the two-body repulsive term 
and the resultant bond length is 3.168\AA and the CSC scheme works efficiently 
(Cf. 2.797\AA without CSC scheme and 2.967\AA  by Gaussian B3LYP/6-31G(d,p)). 
Once we include H 2p state in the calculation, the molecular dimer configuration 
$\alpha=7^\circ$, $\beta=57^\circ$ can be obtained, which is in good agreement 
with the results of Gaussian calculation ($\alpha=6^\circ$ and $\beta=57^\circ$.  
See the inset in Fig.~\ref{Proton}).

Then we simulate the proton transfer in a water system 100H$_2$O + H$_3$O$^+$  
and observe the proton transfer process to happen in a recombination of 
H$_3$O$^{+}$ + H$_2$O $\Rightarrow$ H$_2$O + H$_3$O$^{+}$ (hopping of a proton) 
but not the transfer of an oxonium ion H$_3$O$^{+}$ itself and 
the jumping time is of the order of 1~ps. 
Furthermore, we observe the following elementary process within an order of 0.1~ps;\\
(1) A H$_3$O$^{+}$ ion tries to find a proper one in near-neighboring H$_2$O molecules, \\
(2) a H$_3$O$^{+}$ ion forms a weak bond with one of H$_2$O (forming H$_3$O$^+$+H$_2$O), 
and \\
(3) a proton transfers to that H$_2$O, forming H$_2$O+H$_3$O$^+$. \\
(4) If this H$_2$O is proper one, the new H$_2$O molecule (the former H$_3$O$^{+}$ ion) 
rotates (to form a stable water dimer 2H$_2$O) and change to a stable configuration and, at the same time, \\
(5) a new H$_3$O$^{+}$ ejects a proton H$^+$ to a next H$_2$O.

From these observation, we may conclude that the ASED framework can work properly 
in a wide variety of materials. 
The present system size of water is still too small to use the generalized Lanczos method and 
the generalized shifted COCG method. 
Even so, we have already assured these methods for working in the MD simulation in the present system.

\section{Conclusions}\label{Conclusion}

We have reviewed our recently developed methods of ASED  
based on the generalized H\"uckel approximation 
and the novel linear algebraic algorithm 
for large-scale quantum molecular dynamics simulation. 
The crucial point is that the novel linear algebraic algorithm is applicable to 
systems of several thousand atoms or more with the same accuracy 
of the exact calculation. 
Then we presented examples of the applications to 
the fracture propagation and surface reconstruction in Si and the proton transfer in water.
The water is very sensitive and difficult systems to simulate and 
the successful results suggest that 
our proposed simulation method is promising in a wide variety of systems, 
dense or dilute, solids, liquids or soft materials, and metals or insulators. 
The present MD simulation method was applied also to the problems 
of the formation of helical multishell Au nanowire~\cite{Au-nanowire} 
and the diamond-graphite transformation under applied stress~\cite{diamond-graphite}.

\section*{Acknowledgments}

Numerical calculation was partly carried out 
using the supercomputer facilities of
the Institute for Solid State Physics, University of Tokyo and 
the Research Center for Computational Science, Okazaki. 
Our research progress of large-scale systems and other information can be found 
on the WEB page of ELSES (Extra-Large Scale Electronic Structure calculation) Consortium http://www.elses.jp .

\section*{References}


\end{document}